\PassOptionsToPackage{unicode}{hyperref}
\PassOptionsToPackage{hyphens}{url}
\documentclass[
]{article}
\usepackage{amsmath,amssymb}
\usepackage{iftex}
\ifPDFTeX
  \usepackage[T1]{fontenc}
  \usepackage[utf8]{inputenc}
  \usepackage{textcomp} 
\else 
  \usepackage{unicode-math} 
  \defaultfontfeatures{Scale=MatchLowercase}
  \defaultfontfeatures[\rmfamily]{Ligatures=TeX,Scale=1}
\fi
\usepackage{lmodern}
\ifPDFTeX\else
\fi
\IfFileExists{upquote.sty}{\usepackage{upquote}}{}
\IfFileExists{microtype.sty}{
  \usepackage[]{microtype}
  \UseMicrotypeSet[protrusion]{basicmath} 
}{}
\makeatletter
\@ifundefined{KOMAClassName}{
  \IfFileExists{parskip.sty}{%
    \usepackage{parskip}
  }{
    \setlength{\parindent}{0pt}
    \setlength{\parskip}{6pt plus 2pt minus 1pt}}
}{
  \KOMAoptions{parskip=half}}
\makeatother
\usepackage{xcolor}
\providecommand{\tightlist}{%
  \setlength{\itemsep}{0pt}\setlength{\parskip}{0pt}}
\newtheorem{theorem}{Theorem}
\newtheorem{lemma}{Lemma}

\ifLuaTeX
  \usepackage{selnolig}  
\fi
\IfFileExists{bookmark.sty}{\usepackage{bookmark}}{\usepackage{hyperref}}
\IfFileExists{xurl.sty}{\usepackage{xurl}}{} 
\hypersetup{
  pdftitle={The Geometry of Chi-Square Degrees of Freedom},
  pdfauthor={James Bernhard},
  hidelinks,
  pdfcreator={LaTeX via pandoc}}

\title{The Geometry of Chi-Square Degrees of Freedom}
\author{James Bernhard}
\date{April 23, 2023}

\begin{document}
\maketitle

\begin{abstract}

In this paper, we state and prove a simple geometric interpretation of
the degrees of freedom of a \(\chi^2\) distribution. The general
geometric idea goes back at least to Fisher in the 1920s, but the exact
result does not appear to have been explicitly stated or proved prior to
this paper.

\end{abstract}


What does the term \emph{degrees of freedom} mean in the context of a
\(\chi^2\) distribution? A correct but rather unilluminating answer is
that \emph{degrees of freedom} is the name given to a specific parameter
in the \(\chi^2\) distribution family. But that doesn't explain where
the term comes from or why it makes sense. In this paper, we give a
geometric explanation for the term.

The main result that we prove in this paper is the following.

\begin{theorem}
\label{mainTheorem}
Let $\textbf{X}$ be an $\mathbb{R}^n$-valued normally distributed random vector with mean $\boldsymbol{\mu}$ and variance-covariance operator $C$, and let $W$ be the smallest subspace that contains $\textbf{X}- \boldsymbol{\mu}$ almost surely. Then $\| \textbf{X} \|^2$ has a chi-square distribution if and only if $\boldsymbol{\mu}\in W$ and $C$ is orthogonal projection onto $W$. In this case, the chi-square distribution has $\textrm{\upshape Dim}(W)$ degrees of freedom and noncentrality parameter $\| \boldsymbol{\mu} \|$.
\end{theorem}

In less technical terms, this theorem tells us that
\emph{degrees of freedom} can be thought of as referring to the
dimension of the space that the random variable whose length squared has
a chi-square distribution ``occupies.'\,'

Although he didn't use the exact language of this theorem, Fisher ---
who coined the term \emph{degrees of freedom} and was instrumental in
developing the theory of \(\chi^2\) distributions --- was himself
thinking along these lines, as the following quotation (originally from
{[}1{]} and reprinted in {[}2{]}) shows:

\begin{quote}
In the writer's thought, though not very explicitly in this paper, the mathematical distribution given by tables of $\chi^2$ was that of the sum of $n$ squares of variates normally and independently distributed about zero with unit variance. $\chi^2$ in fact was the square of the distance of a random point from the centre of a homogeneous normal distribution in $n$ dimensions. The number of dimensions, however, would be reduced by unity for every restriction upon deviations between expectation and observation, and it appeared that the inconsistencies in the literature could be straightened out if account were taken of the true number of degrees of freedom in which observation and expectation might in reality differ. 
\end{quote}

Given Fisher's point of view and how straightforward the geometric
explanation is, it may seem surprising that (as far as the author has
been able to determine) prior to this paper, no such explanation of the
term \emph{degrees of freedom} has been written out explicitly. This may
be in part because instead of thinking of the most common \(\chi^2\)
distributed random variables in statistics as arising as
\(\| \textbf{X} \|^2\) where \(\textbf{X}\sim N(\boldsymbol{\mu}, C)\),
people have mostly considered them as arising as
\(\textbf{X} \cdot A(\textbf{X})\) where
\(\textbf{X}\sim N(\boldsymbol{\mu}, I)\). Cochran's theorem tells us
that, for positive definite, self-adjoint \(A\), the random variable
\(\textbf{X} \cdot A(\textbf{X})\) has a chi-square distribution if and
only if \(A^2 = A\). But in this case, \[
    \textbf{X} \cdot A(\textbf{X}) = \| A(\textbf{X}) \|^2,
\] so it is a matter of preference whether to consider such a random
variable as coming from a quadratic form applied to a normally
distributed random vector \(\textbf{X}\) with variance-covariance
operator \(I\), or as coming from the squared length of of a normally
distributed random vector \(A(\textbf{X})\) with a less restricted
variance-covariance operator. Traditionally the first approach has been
used, but that obscures the geometry. In this paper, we explore the
second approach.

To help us prove Theorem \ref{mainTheorem}, we first give a geometric
interpretation of the variance-covariance operator of a random vector.

\section{Geometry of the variance-covariance operator}

The first step towards understanding degrees of freedom geometrically is
to understand the variance-covariance operator of a random vector
geometrically. In this section, we describe its kernel and its image.

The following lemma is well-known, but we include it for completeness of
the discussion.

\begin{lemma}
\label{lemma-dispersionKernel}
Let $\textbf{X}$ be an $\mathbb{R}^n$-valued random vector with mean $\boldsymbol{\mu}$ and variance-covariance operator $C$. Then
$$
    \textrm{\upshape Ker}(C) = \{ \textbf{\upshape v}\in \mathbb{R}^n|(\textbf{X}- \boldsymbol{\mu}) \cdot \textbf{\upshape v} = 0 \text{\upshape\ almost surely})\}.
$$
\end{lemma}

\noindent\emph{Proof.} Let
\(\textbf{\upshape v}, \textbf{\upshape w}\in \mathbb{R}^n\). Then
\begin{align*}
C(\textbf{\upshape v}) \cdot \textbf{\upshape w} &= \textbf{\upshape v} \cdot C(\textbf{\upshape w})\\
&= \textrm{\upshape Cov}(\textbf{X} \cdot \textbf{\upshape v}, \textbf{X} \cdot \textbf{\upshape w})\\
&= E(((\textbf{X}-\boldsymbol{\mu}) \cdot \textbf{\upshape v})((\textbf{X}-\boldsymbol{\mu}) \cdot \textbf{\upshape w})),
\end{align*} and \(\textbf{\upshape v}\in \ker{C}\) if and only if this
equals \(0\) for all \(\textbf{\upshape w}\in \mathbb{R}^n\).

Suppose \(\textbf{\upshape v}\in \ker{C}\). Then the above equals \(0\)
for all \(\textbf{\upshape w}\in \mathbb{R}^n\), so in particular \[
E( ((\textbf{X}- \boldsymbol{\mu}) \cdot \textbf{\upshape v})^2 ) = 0,
\] which implies that
\((\textbf{X}- \boldsymbol{\mu}) \cdot \textbf{\upshape v} = 0\) almost
surely.

On the other hand, if
\((\textbf{X}- \boldsymbol{\mu}) \cdot \textbf{\upshape v} = 0\) almost
surely, then \[
P( ((\textbf{X}- \boldsymbol{\mu}) \cdot \textbf{\upshape v})((\textbf{X}- \boldsymbol{\mu}) \cdot \textbf{\upshape w}) = 0) = 1\text{ for all } \textbf{\upshape w}\in \mathbb{R}^n,
\] so \(\textbf{\upshape v}\in \ker{C}\). \ensuremath{\diamond}

With this, we can now describe the image of a random vector's
variance-covariance operator geometrically. The following result,
although presumably fairly widely known, does not seem to appear
anywhere in the published literature.

\begin{theorem}
Let $\textbf{X}$ be an $\mathbb{R}^n$-valued random vector with mean $\boldsymbol{\mu}$ and variance-covariance operator $C$. Then $\textrm{\upshape Image}(C)$ is the smallest subspace that contains $\textbf{X}- \boldsymbol{\mu}$ almost surely.
\end{theorem}

\noindent\emph{Proof.} Let \(W_X\) denote the smallest subspace that
contains \(\textbf{X}- \boldsymbol{\mu}\) almost surely, which is the
intersection of all subspaces with this property.

To see that \(W_X \subseteq \textrm{\upshape Image}(C)\), note that
since \(C\) is self-adjoint, then
\(\textrm{\upshape Image}(C) = \textrm{\upshape Ker}(C)^\perp\). Let
\(\{ \textbf{\upshape b}_1, \ldots, \textbf{\upshape b}_k \}\) be an
orthonormal basis for \(\textrm{\upshape Ker}(C)\). Then \begin{align*}
    P(\textbf{X}- \boldsymbol{\mu}\in \textrm{\upshape Image}(C)) &= P(\textbf{X}- \boldsymbol{\mu}\in \textrm{\upshape Ker}(C)^\perp)\\
    &= P((\textbf{X}- \boldsymbol{\mu}) \cdot \textbf{\upshape b}_i = 0\text{ for all $i = 1, \ldots, k$})\\
    &= 1 - P((\textbf{X}- \boldsymbol{\mu}) \cdot \textbf{\upshape b}_i \neq 0\text{ for some $i = 1, \ldots, k$})\\
    &\geq 1 - \sum_{i=1}^k P((\textbf{X}- \boldsymbol{\mu}) \cdot \textbf{\upshape b}_i \neq 0)\\
    &= 1,
\end{align*} since each of the events being considered has probability
\(0\) by Lemma \ref{lemma-dispersionKernel}. So
\(\textbf{X}- \boldsymbol{\mu}\in \textrm{\upshape Image}(C)\text{\upshape\ almost surely}\),
which means that \(W_X \subseteq \textrm{\upshape Image}(C)\).

To see that \(\textrm{\upshape Image}(C) \subseteq W_X\), let
\(W \subseteq \mathbb{R}^n\) be a subspace with
\(\textbf{X}- \boldsymbol{\mu}\in W\text{\upshape\ almost surely}\). If
\(\textbf{\upshape v}\in W^\perp\), then
\((\textbf{X}- \boldsymbol{\mu}) \cdot \textbf{\upshape v} = 0\text{\upshape\ almost surely}\),
so Lemma \ref{lemma-dispersionKernel} implies that
\(\textbf{\upshape v}\in \textrm{\upshape Ker}(C) = \textrm{\upshape Image}(C)^\perp\).
This means that \(W^\perp \subseteq \textrm{\upshape Image}(C)^\perp\),
so \(\textrm{\upshape Image}(C) \subseteq W\).

Since \(\textrm{\upshape Image}(C) \subseteq W\) for all subspaces \(W\)
that contain \(\textbf{X}- \boldsymbol{\mu}\) almost surely, then
\(\textrm{\upshape Image}(C) \subseteq W_X\), their intersection.
\ensuremath{\diamond}

Given how useful the smallest subspace that contains
\(\textbf{X}- \boldsymbol{\mu}\) almost surely is, it is tempting to
give such a subspace a name. However, such additional terminology is
unnecessary, since this can now be viewed as just another way to think
about the image of the variance-covariance matrix of \(\textbf{X}\).

Now that we understand the variance-covariance operator geometrically,
we can prove the main result.

\hypertarget{proof-of-theorem}{%
\section{\texorpdfstring{Proof of Theorem
\ref{mainTheorem}}{Proof of Theorem }}\label{proof-of-theorem}}

To prove Theorem \ref{mainTheorem}, we first prove the following lemma,
which establishes algebraically the core of the theorem. We will then
use the results of the previous section to interpret this lemma
geometrically and prove Theorem \ref{mainTheorem}.

\begin{lemma}
Let $\textbf{X}\sim N(\boldsymbol{\mu}, C)$. Then $\| \textbf{X} \|^2$ has a chi-square distribution if and only if $C^2 = C$ and $C(\boldsymbol{\mu}) = \boldsymbol{\mu}$. In this case, $\| \textbf{X} \|^2 \sim \chi^2(\textrm{\upshape Rank}(C), \| \boldsymbol{\mu} \|)$.
\end{lemma}

\noindent\emph{Proof.} The moment-generating function of
\(\| \textbf{X} \|^2\) is (see {[}3{]}, pages 7, 13, and 23, although
page 23 has a minor typographical error): \[
    M_1(t) = \frac{\exp\left(t\boldsymbol{\mu} \cdot (I-2tC)^{-1}(\boldsymbol{\mu})\right)}{\sqrt{\det(I-2tC)}},
\] for all suitably small \(t\).

To calculate this more explicitly, let
\(\mathcal{B}= \{ \textbf{\upshape b}_1, \ldots, \textbf{\upshape b}_n \}\)
be an orthonormal eigenbasis for \(\mathbb{R}^n\) relative to \(C\) with
eigenvalues \(\lambda_1, \ldots, \lambda_n\) respectively. Also, suppose
that \(\boldsymbol{\mu}= \sum_{i=1}^n \mu_i \textbf{\upshape b}_i\).
Then we have \[
M_1(t) = \exp\left(t \sum_{i=1}^n (1-2t\lambda_i)^{-1}\mu_i^2\right) \left( \prod_{i=1}^n 1-2t\lambda_i \right)^{-1/2}.
\] This means that the cumulant-generating function of
\(\| \textbf{X} \|^2\) is \begin{align*}
\log(M_1(t)) &= \sum_{i=1}^n t(1-2t\lambda_i)^{-1}\mu_i^2 -\frac{1}{2} \log(1-2t\lambda_i)\\
&= \sum_{i=1}^n t \left( \sum_{j=0}^\infty (2t\lambda_i)^j \right) \mu_i^2 - \frac{1}{2} \left( \sum_{k=1}^\infty \frac{-(2t\lambda_i)^k}{k} \right)\\
&= \sum_{j=1}^\infty 2^{j-1} j! \left( \sum_{i=1}^n \lambda_i^{j-1}(\frac{\lambda_i}{j} + \mu_i^2)\right) \frac{t^j}{j!}. 
\end{align*} That is, for \(j = 1, 2, \ldots\), the \(j\)-th cumulant of
\(\| \textbf{X} \|^2\) is \begin{align}
    2^{j-1} (j-1)! \left( \sum_{i=1}^n \lambda_i^j + j \sum_{i=1}^n \lambda_i^{j-1}\mu_i^2) \right).\label{cumulants}
\end{align}

To prove the \emph{if} part of the theorem, suppose that \(C^2 = C\) and
\(C(\boldsymbol{\mu}) = \boldsymbol{\mu}\). The first of these implies
that \(\lambda_i = 0\text{ or }1\) for all \(i = 1, \ldots, n\), so the
above expression becomes \[
    \begin{cases}
    \displaystyle
    2^{j-1} (j-1)! \left( \sum_{i=1}^n \lambda_i + j \sum_{i=1}^n \mu_i^2 \right) &\text{for $j=1$}\\
    \displaystyle
    2^{j-1} (j-1)! \left( \sum_{i=1}^n \lambda_i + j \sum_{i=1}^n \lambda_i\mu_i^2 \right) &\text{for $j=2, 3, \ldots$}.
    \end{cases}
\] Since \(\boldsymbol{\mu}= C(\boldsymbol{\mu})\), then
\(\displaystyle \sum_{i=1}^n \lambda_i \mu_i^2 = \boldsymbol{\mu} \cdot C(\boldsymbol{\mu}) = \| \boldsymbol{\mu} \|^2 = \sum_{i=1}^n \mu_i^2\).
Also, since \(C^2 = C\) and \(C\) is self-adjoint, then
\(\sum_{i=1}^n \lambda_i = \textrm{\upshape Rank}(C)\), so these two
cases can be combined into a single expression: \[
    2^{j-1} (j-1)! \left( \textrm{\upshape Rank}(C) + j \| \boldsymbol{\mu} \|^2 \right)\quad\text{for $j=1, 2, 3, \ldots$}.
\] These are the cumulants of a
\(\chi^2(\textrm{\upshape Rank}(C), \| \boldsymbol{\mu} \|)\)
distribution, which proves the \emph{if} part of the theorem.

The \emph{only if} part of the theorem is more subtle than it might
first appear. This is discussed carefully in {[}4{]}, and we use here a
modified version of the proof of a similar result given there.

Suppose that \(\| \textbf{X} \|^2 \sim \chi^2(r, \nu)\) for some
\(r, \nu\). Then the moment-generating function of
\(\| \textbf{X} \|^2\) is \[
    M_2(t) = \frac{e^{t\nu^2/(1-2t)}}{(1-2t)^{r/2}}
\] for all suitably small \(t\). This implies that the
cumulant-generating function of \(\| \textbf{X} \|^2\) is \begin{align*}
\log(M_2(t)) &= -\frac{r}{2} \log(1-2t) + \frac{t\nu^2}{1-2t}\\
&= -\frac{r}{2} \left( \sum_{j=1}^\infty \frac{-(2t)^n}{n} \right) + t\nu^2 \left( \sum_{k=0}^\infty (2t)^k \right)\\
&= \sum_{j=1}^\infty 2^{j-1} (j-1)! (r+ j\nu^2) \frac{t^j}{j!}.
\end{align*} This means that the cumulants of \(\| \textbf{X} \|^2\) are
\[
    2^{j-1} (j-1)! (r+ j\nu^2)\quad\text{for $j=1, 2, \ldots$}.
\] Setting this equal to the expression \eqref{cumulants} obtained above
for the cumulants of \(\| \textbf{X} \|^2\), we get that \begin{align*}
    \frac{r}{j} + \nu^2 &= \sum_{i=1}^n \lambda_i^{j-1}(\frac{\lambda_i}{j} + \mu_i^2)\quad\text{for $j=1, 2, \ldots$}.
\end{align*} It is useful to split this sum into two pieces:
\begin{align}
    \frac{r}{j} + \nu^2 &= \sum_{i|\lambda_i = 1} (\frac{1}{j} + \mu_i^2) + \sum_{i|\lambda_i < 1} \lambda_i^{j-1}(\frac{\lambda_i}{j} + \mu_i^2)\quad\text{for $j=1, 2, \ldots$}.\label{cumulantEquation}
\end{align} Also, since \(C\) is positive semidefinite, then
\(\lambda_i \geq 0\) for all \(i=1, \ldots, n\).

The left side of \eqref{cumulantEquation} is bounded as
\(j \to \infty\), so the right side must be as well, which implies that
\(\lambda_i \leq 1\) for all \(i=1, \ldots, n\).

Using this, taking the limit as \(j \to \infty\) on both sides gives
\begin{align}
    \nu^2 &= \sum_{i|\lambda_i = 1} \mu_i^2.\label{ncp}
\end{align} Putting this into the previous equation and simplifying
gives \[
    r= \sum_{i|\lambda_i = 1} 1 + \sum_{i|\lambda_i < 1} \lambda_i^{j-1}(\lambda_i + j\mu_i^2)\quad\text{for $j=1, 2, \ldots$}.
\] Taking the limit as \(j \to \infty\) of both sides, we get
\begin{align}
    r&= \sum_{i|\lambda_i = 1} 1.\label{df}
\end{align} Putting this into the previous equation and simplifying
gives \[
    \sum_{i|\lambda_i < 1} \lambda_i^{j-1}(\lambda_i + j\mu_i^2) = 0\quad\text{for $j=1, 2, \ldots$}.
\] From the \(j=1\) case, this implies that \[
    \sum_{i|\lambda_i < 1} \lambda_i + \mu_i^2 = 0.
\] Since \(\lambda_i \geq 0\) and \(\mu_i^2 \geq 0\) for all
\(i=1, \ldots, n\), this means that \[
    \lambda_i = 0\text{ and } \mu_i = 0\quad\text{whenever $\lambda_i < 1$}.
\] Putting this all together, we have found that \(0\) and \(1\) are the
only eigenvalues of \(C\), so \(C^2 = C\). Also, since \(\mu_i = 0\)
whenever \(\lambda_i = 0\) and since \(\lambda_i = 1\) whenever
\(\lambda_i \neq 0\), then \(C(\boldsymbol{\mu}) = \boldsymbol{\mu}\).

Also, by \eqref{df} and since all the nonzero eigenvalues are \(1\),
then \(r\) is the number of nonzero eigenvalues. Since \(C\) is
self-adjoint and therefore diagonalizable, then this sum equals
\(\textrm{\upshape Rank}(C)\).

By \eqref{ncp} and since \(\mu_i = 0\) whenever \(\lambda_i = 0\), then
\(\displaystyle \nu^2 = \sum_{i=1}^n \mu_2 = \| \boldsymbol{\mu} \|^2\).
Since both sides are nonnegative, then \(\nu= \| \boldsymbol{\mu} \|\),
which completes the proof of the theorem.~\ensuremath{\diamond}

Combining this with the previous section's results, we arrive at the
main result. \setcounter{theorem}{0}

\begin{theorem}
Let $\textbf{X}\sim N(\boldsymbol{\mu}, C)$ with $C$ self-adjoint and positive semidefinite, and let $W$ be the smallest subspace that almost surely contains $\textbf{X}- \boldsymbol{\mu}$. Then $\| \textbf{X} \|^2$ has a chi-square distribution if and only if $C= \textrm{\upshape Proj}_{W}$ and $\boldsymbol{\mu}\in W$. In this case, $\| \textbf{X} \|^2 \sim \chi^2(\textrm{\upshape Dim}(W), \| \boldsymbol{\mu} \|)$.
\end{theorem}

\noindent\emph{Proof.} By the previous theorem, \(\| \textbf{X} \|^2\)
has a chi-square distribution if and only if \(C^2 = C\) and
\(C(\boldsymbol{\mu}) = \boldsymbol{\mu}\). Since \(C\) is self-adjoint,
then \(C^2 = C\) if and only if \(C\) is an orthogonal projection onto
its image, which by the previous section is \(W\). Also, given that
\(C\) is an orthogonal projection onto \(W\), then
\(C(\boldsymbol{\mu}) = \boldsymbol{\mu}\) if and only if
\(\boldsymbol{\mu}\in W\). \ensuremath{\diamond}

This theorem gives a geometric explanation of the term
\emph{degrees of freedom} of a chi-square distribution. Chi-square
distributions arise as the squared length of a normally distributed
random vector \(\textbf{X}\). This theorem tells us the conditions under
which this happens, and that when it happens, the number of degrees of
freedom of \(\| \textbf{X} \|^2\) is the dimension of the smallest
subspace that almost surely contains \(\textbf{X}- \boldsymbol{\mu}\)
(or, equivalently, the dimension of the image of the variance-covariance
operator of \(\textbf{X}\)).

\hypertarget{bibliography}{%
\section{Bibliography}\label{bibliography}}

\begin{enumerate}
\def\labelenumi{\arabic{enumi}.}
\tightlist
\item
  R.A. Fisher,
  \emph{On the Interpretation of $\chi^2$ from Contingency Tables, and the Calculation of P},
  Journal of the Royal Statistical Society, \textbf{85} (1922), no. 3,
  87-94.
\item
  R.A. Fisher, \emph{Contributions to Mathematical Statistics}, John
  Wiley \& Sons, New York, 1950.
\item
  Issie Scarowsky, \emph{Quadratic Forms in Normal Variables},
  unpublished master's thesis, McGill University, Department of
  Mathematics.
\item
  Michael F. Driscoll,
  \emph{An Improved Result Relating Quadratic Forms and Chi-Square Distributions},
  The American Statistician, \textbf{53} (1999) No.~3, 273-275.
\end{enumerate}

\end{document}